\documentclass[twocolumn,showpacs,preprintnumbers,amsmath,amssymb]{revtex4}

\usepackage{graphicx}
\usepackage{dcolumn}
\usepackage{bm}
\usepackage{color}
\usepackage{amsmath}
\usepackage{amssymb}
\usepackage{dsfont}
\usepackage{color}
\usepackage{calc}
\usepackage{hyperref}
\usepackage{multirow}
\usepackage{subfigure}
\usepackage{float}
\usepackage[normalem]{ulem}
\usepackage{natmove}

\def\be{\begin{equation}}
\def\ee{\end{equation}}
\def\bea{\begin{eqnarray}}
\def\eea{\end{eqnarray}}

\graphicspath{{./figures/}}

\begin{document}

\title{Quantum-phase-field concept of matter: Emergent gravity in the dynamic universe}

\author{I.~Steinbach}
 \email{ingo.steinbach@rub.de}

\affiliation{ Ruhr-University Bochum, ICAMS, Universitaetsstrasse 150, 44801 Bochum, Germany}

\pacs{04.20.Cv, 04.50.Kd, 05.70.Fh}


\date{\today}

\begin{abstract}
A monistic framework is set up where energy is the only fundamental substance. Different states of energy are ordered by a set of scalar qunatum-phase-fields. The dual elements of matter, mass and space, are described as volume- and gradient-energy contributions of the set of fields, respectively. Time and space are formulated as background-independent dynamic variables. The  evolution equations of the body of the universe are derived from first principles of thermodynamics. Gravitational interaction emerges from quantum fluctuations in finite space. Application to a large number of fields predicts scale separation in space and repulsive action of masses distant beyond a marginal distance. The predicted marginal distance is compared to the size of the voids in the observable universe.
\end{abstract}


\maketitle
{Corresponding to the published version in: Zeitschrift f\"ur Naturforschung A; 72, 51-58 (2017); DOI 10.1515/zna-2016-0270, with several formal corrections.}

\section{Introduction}\label{Intro}

"Several recent results suggest that the field equations of gravity have the same conceptual status as the equations of, say, elasticity or fluid mechanics, making gravity an emergent phenomenon" starts the review of Padmanabhan and Padmanabhan on the cosmological constant problem \cite{Pad14}. This point of view relates to the holographic principle \cite{tHooft1993,Verlinde2011,Bousso2002} which treats gravity as an `entropic force' derived from the laws of thermodynamics. Even more radical approaches are given by the `causal sets' of Sorkin \cite{Sorkin1991} and Rovelli's `loop-space representation of quantum general-relativity' \cite{Rov1990}, which treat space-time as fundamentally discrete. I will adopt from the latter that there is no fundamental multi-dimensional continuous space-time, but a discrete set of fields. From the first that thermodynamics shall be the fundament of our understanding of the world. 

The concept is based on a formalism which is well established in condensed matter physics, the so-called phase-field theory (for review see \cite{boettinger_phase-field_2002,Steinbach2013}). It is applied to investigate pattern formation in  mesoscopic bodies where no length scale is given. Mesoscopic in this context means `large compared to elementary particles or atoms' and `small compared to the size of the body'. Then the scale of a typical pattern is treated emergent from interactions between different elements of the body under investigation. The general idea of the phase-field theory is to combine energetics of surfaces with volume thermodynamics. It is interesting to note that it thereby inherits the basic elements of the holographic principle which relates the entropy of a volume in space-time to the entropy at the surface of this volume. In the phase-field theory, the competition of the free energy of volume and surface drives the evolution of the system under consideration. I will start out from first principles of energy conservation and entropy production in the general form of \cite{Lieb1999}. Energy is the only fundamental substance. `Fundamental substance' in this context means `a thing-in-itself, regardless of its appearance' \cite{Langton_2001}. There will be positive and negative contributions to the total energy $H$. They have to be balanced to zero since there is no evidence, neither fundamental nor empirical, for a source where the energy could come from: $H~=~<w|\hat H|w>~=~0$. I will call this the `principle of neutrality'. Compare also the theory of Wheeler and DeWitt \cite{DeWitt1967} which is, however, based on a fundamentally different framework in relativistic quantum-mechanics. The Hamiltonian $\hat H$ will be expanded as a function of the quantum-phase-fields $\{\phi_I\}$,  $I=1...N$, and their gradients. The wave function $|w>$ will be treated explicitly in the limiting case of quasi-stationary elementary masses. The time dependence of the Hamiltonian and the wave function is governed by relaxational dynamics of the fields according to the demand of entropy production. Here I will treat  interaction of neutral matter only. Additional quantum numbers like charge and color may be added to the concept later.

\section{Basic considerations}\label{Basic}

The concept is based on the following statements:

\begin{itemize}
\item The first and the second laws of thermodynamics apply.
\item Energy is fundamental and the principle of neutrality applies, i.e. the total energy of the  universe is zero.
\item There is the possibility that energy separates into two or more different states.
\item Different states of energy can be ordered by a set of $N$ dimensionless quantum-phase-fields $\{\phi_I\}$, $I=1...N$. The fields have normalized bounds $0~\le~\phi_I~\le~1$.
\item The system formed by the set of fields is closed in itself:
\be\label{sum}
\sum_{I=1}^N~\phi_I~=~1.
\ee
\item Two 1-dimensional metrices, evaluating distances between states of energy, define space and time as dynamic variables.
\item Planck's constant $h$, the velocity $c$ and the kinetic constant $m_0$ with the dimension of mass are universal.
\item Energy and mass are proportional with the constant $c^2$.
\end{itemize}

One component of the set of fields $\{\phi_I\}$ is considered as an `order parameter' in the sense of Landau \cite{Landau1959}, characterizing the state of a subsystem of the body under consideration. The `$0$' value of the field $\phi_I$ denotes that this  state is not existing. The value `$1$' means that this state is the only one existing. Intermediate values mean coexistence of several  states. There is obviously a trivial solution of (\ref{sum}): $\phi_I=\frac1N$. For this solution no `shape' can be distinguished. It is one possible homogeneous state of the body. "There is, however, no reason to suppose that $[...]$ the body $[...]$ will be homogeneous." (\cite{Landau1959} page 251). We shall allow phase separation by the demand of entropy production. Phase separation requires the introduction of a metric which allows to distinguish between objects (parts of the body): `space'. Now that we have already two fundamentally different states of the body, the homogeneous state and the phase-separated state, we need a second, topologically different metric to distinguish these states: `time'. Both coordinates, space and time are dynamic,  dependent only on the actual state of the body. They are background independent having no `global' meaning in the sense that they would be independent of the observer. For general considerations about a dynamical universe, see Barbour's dynamical theory \cite{Barbour1974}. For discussions about the `arrow of time', see \cite{Kiefer1995}.

\section {Variational framework}\label{Variational}

The concept is based on the variational framework of field theory \cite{Kiselev}. The energy functional $\hat H$ is defined by the integral over the energy density $\hat h$ as a function of the fields $\{\phi_I\}$ with a characteristic length $\eta$, to be determined:

\be\label{density}
\hat H = \eta \sum_{I=1}^N \int_0^1 d \phi_I \hat h(\{\phi_I\}).
\ee

The functional $\hat H$ has the dimension of energy and the density $\hat h$ has the dimension of force. The functional (\ref{density}) shall be expanded in the distances $\tilde s_I$

\bea\label{density_space}
\hat H &=& \eta \sum_{I=1}^N \int_{-\infty}^\infty d\tilde s_I \frac {\partial \phi_I}{\partial \tilde s_I} \hat h(\{\phi_I\}) \\
&=& \sum_{I=1}^N \int_{-\infty}^\infty ds_I \hat h(\{\phi_I\}),
\eea
where distances are renormalized according to

\be\label{renorm}
s_I=\eta\tilde s_I \frac {\partial \phi_I}{\partial \tilde s_I}.
\ee

For readability, I will omit the field index $I$ of the distances in the following. The individual fields are functions in distances in space and time $\phi_I=\phi_I(s,t)$. They will be embedded into a higher dimensional mathematical space in section \ref{MI}. The time evolution of one field is determined by relaxational dynamics with a relaxation constant $\tilde \tau \propto m_0$:

\bea\label{statistical_motion}
 \tilde \tau  \frac\partial{\partial t} \phi_I = -\frac{\delta}{\delta\phi_I} \int_{0}^{+\infty}dt <w| \hat H|w>.
\eea

I use the standard form of the Ginzburg-Landau functional, or Hamiltonian, $\hat H$ in 2-dimensional Minkowski notation, the time derivative accounting for dissipation. 

\bea\label{functional_H}
 \hat H =  \sum_{I=1}^N \int_{-\infty}^{+\infty}ds \frac {4 U\eta}{\pi^2} \;\;\;\;\;\;\;\;\;\;\;\;\;\;  \\ \nonumber \{(\frac \partial{\partial s} \phi_I)^2 - \frac1{c^2}(\frac \partial{\partial t} \phi_I)^2 + \frac {\pi^2} {\eta^2} |\phi_I (1-\phi_I)|\},
\eea
where $U$ is a positive energy quantum to be associated with massive energy. Note that the special analytical form of this expansion is selectable as long as isotropy in space-time is guaranteed and the dual elements of gradient and volume contributions are normalized to observable physical quantities, see eq. (\ref{mass_energy}) and (\ref{EC}) below. 

\section{Quasi-static solution}\label{OSL}

Now I will formally derive the individual constituents of the concept related to known physical entities in mechanics. I will only treat the quasi-static limit where the dynamics of the wave function $|w>$ and the dynamics of the fields $\phi_I$ decouple. This means that the fields are kept static for the quantum solution on the one hand. The quantum solution on the other hand determines the energetics of the fields. The expectation value of the energy functional (\ref{functional_H}) has three formally different contributions if the differential operators  $\frac {\partial}{\partial s}$ and  $\frac {\partial}{\partial t}$ are applied to the wave function $|w>$ or the fields $\phi_I$ respectively.

Applying the differential operators to the fields and using  the normalization of the wave function $<w|w>=1$ yields the force $u_{I}$ related to the gradient of the fields $I$:

\bea\label{u_interface}
u_{I} =  \frac {4U\eta}{\pi^2}\left[(\frac {\partial  \phi_I}{\partial s})^2 -\frac1{c^2}(\frac {\partial \phi_I}{\partial t})^2 +\frac{\pi^2}{\eta^2} |\phi_I(1-\phi_I)|\right].
\eea

The mixed contribution describes the correlation between the fields and the wave function and shall be set to $0$ in the quasi-static limit:

\bea\label{acceleration}
\phi_I \frac {4U\eta}{\pi^2} \Bigm[ \frac {\partial \phi_I}{\partial s} <w|\frac {\partial}{\partial s}|w>   \\
 \nonumber - \frac1{c^2}\frac {\partial \phi_I }{\partial t}<w|\frac {\partial}{\partial t}|w> \Bigm]  &=& 0.
\eea

The force $e_I$ related to the volume of field $I$ is defined:

\be\label{energydensity}
e_I|_{\phi_I=1} =\frac{4U\eta}{\pi^2} \phi_I^2 <w|\frac {\partial^2}{\partial s^2}-\frac1{c^2}\frac {\partial^2}{\partial t^2}|w>.
\ee

Next we need to elaborate the structure of the fields. I do this for the special case of $N=2$ in a linear setting with periodic boundary conditions, for simplicity. The general case is a straight forward extension which cannot be solved analytically, however. For the analytic solvability it is also convenient to replace the coupling function $\phi_I^2$ in (\ref{energydensity}) by the function $m(\phi)=\frac \pi 2 \{(2\phi-1)\sqrt {\phi(1-\phi)}+\frac 12 \arcsin {2\phi-1}\}$ which is monotonous between the states $0$ and $1$ and has the normalization to $0$ and $1$ for these states. Differences in both coupling functions become irrelevant in the sharp interface limit $\eta\rightarrow0$ to be investigated here. For $N=2$ and $\phi_1 = 1-\phi_2=\phi$ the equation of motion (\ref{statistical_motion}) read, with $\Delta e = e_1 - e_2$, $m_{\phi}=\frac {\partial m}{\partial \phi}$ and $\tau=\frac {\pi^2}8 \tilde\tau$:

\bea\label{min}
\tau\frac\partial{\partial t}\phi = \tau v \frac\partial{\partial s}\phi = \\
\nonumber  U [\eta \frac {\partial^2\phi}{\partial s^2}(1-\frac{{v}^2}{c^2}) &+& \frac{\pi^2}\eta (\phi-\frac12)] + m_{\phi}\Delta e.
\eea

I have transformed the time derivative of the field $\frac {\partial}{\partial t}\phi$ into the moving frame with velocity $v$, $\frac {\partial}{\partial t}= v \frac {\partial}{\partial s}$ and used the Euler-Lagrange relation
\be\label{EulerLagrange}
\frac{\delta}{\delta\phi} \int_{-\infty}^{+\infty} ds \int_{0}^{+\infty}dt \rightarrow \frac \partial{\partial\phi}-\frac \partial{\partial t}\frac \partial{\partial\phi_t}-\frac \partial{\partial s}\frac \partial{\partial\phi_s}.
\ee

\begin{figure}[ht]
 \centering

\includegraphics[width=8.0cm]{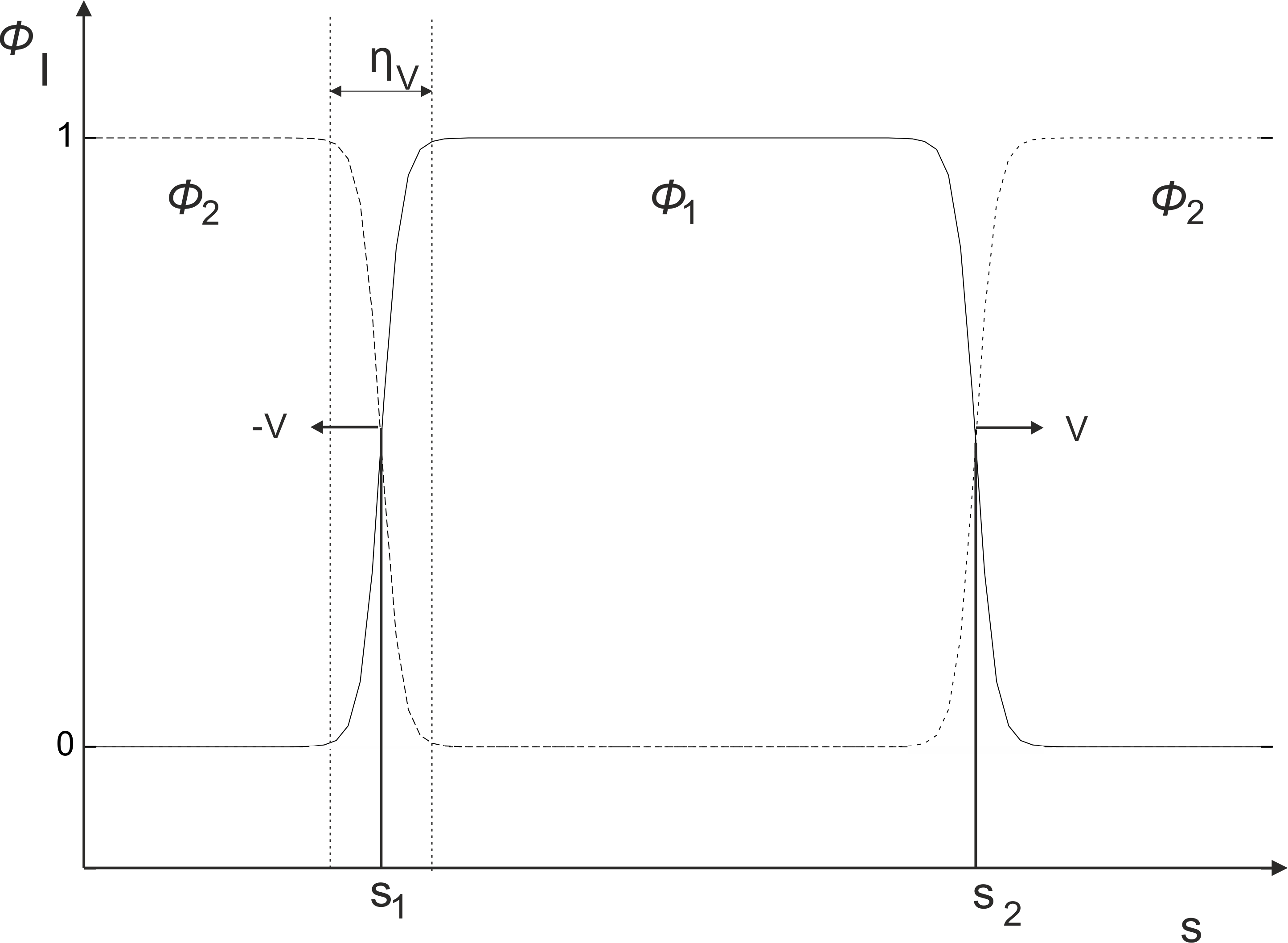}
\caption{Travelling wave solution for two fields in linear arrangement and with periodic boundary conditions. }\label{chain}
\end{figure}

The contributions of (\ref{min}) proportional to $U$ dictate from their divergence in the limit $\eta\rightarrow 0$ the special solution for the field, which is the well known `solution of a traveling wave', or `traveling wave solution' (see appendix of \cite{Steinbach2009a}). We find, besides the trivial solution $\phi(s,t) \equiv 0$, the primitive solution ($s_1<s_2$)
\bea
\label{travellingWave}
\phi =\begin{cases} 0 & \mbox{for}\;\;s<s_1-vt-\frac{\eta_v}2\\ 
\frac 12&+ \frac 12 sin(\frac{\pi(s-s_1+vt)}{\eta_v})\\
&\mbox{for}\;\; s_1-vt-\frac{\eta_v}2\le s < s_1-vt+\frac{\eta_v}2 \\
 1 & \mbox{for}\;\; s_1-vt+\frac{\eta_v}2\le s < s_2+vt-\frac{\eta_v}2 \\
\frac 12&- \frac 12 sin(\frac{\pi(s-s_2-vt)}{\eta_v})\\
&\mbox{for}\;\; s_2+vt-\frac{\eta_v}2\le s < s_2+vt+\frac{\eta_v}2 \\
0 & \mbox{for}\;\;s\ge s_2+vt+\frac{\eta_v}2
\end{cases}
\eea
where $\eta_v= \eta\sqrt{1-\frac{v^2}{c^2}}$ is the effective size of the transition region, or junction, between the fields which I will call `particle' in the following. It is a highly localized state of positive energy (see eq. (\ref{mass_energy} below). The size $\eta_v=\eta_v(v)$ is a function of velocity. $s_1$ and $s_2$ are the spatial coordinates of the particles in the quasi-static picture related to the distance $\Omega~=~|s_1-s_2|$. Figure~\ref{chain} depicts the solution for two fields where the particles travel with velocity $v$ and have finite extension $\eta_v$. It will be treated in the ´sharp interface limit' $\eta\rightarrow 0$ where its extension is negligible compared to distances between objects, but finite as discussed in chapter \ref{speculations}.

Continuing the analysis of (\ref{travellingWave}), one easily proves
\bea\label{derivatives}
\frac\partial{\partial s} \phi |_\text{left}&=& - \frac\partial{\partial s} \phi |_\text{right}= \frac \pi {\eta_v}\sqrt{\phi(1-\phi)},\\
\frac{\partial^2}{\partial s^2} \phi |_\text{left}&=& \frac{\partial^2}{\partial s^2} \phi |_\text{right}= \frac{\pi^2}{\eta_v^2}(\frac12-\phi),\;\;\;
\eea
and we find, as a check for consistency, the energy of two particles from the integral
\be\label{mass_energy}
 \int_{-\infty}^{+\infty}ds \frac{4U}{\pi^2}\left[\eta_v(\frac {\partial \phi}{\partial s} )^2+\frac{\pi^2}{\eta_v}| \phi(1-\phi)|\right]=2 U.
\ee

\subsection{ Volume energy of the fields} \label{energy_density}

From the solution (\ref{travellingWave}), we see that the field in the sharp interface limit forms a one-dimensional box with fixed walls and size $\Omega_I$ for field $I$. According to Casimir \cite{Casimir1948}, we have to compare quantum fluctuations in the box with discrete spectrum $p$ and frequency $\omega_p= \frac {\pi c p }{2\Omega_I}$ to a continuous spectrum. This yields the negative energy $E_I$ of the field $I$:
\be\label{EC}
E_I = \alpha \frac {h c}{4\Omega_I} \left[ \sum_{p=1}^\infty p - \int_1^{\infty} p dp \right]\\ = - \alpha \frac {h c}{48\Omega_I},
\ee
where $\alpha$ is a positive, dimensionless coupling coefficient to be determined. I have used Euler--MacLaurin formula in the limit $\epsilon \rightarrow 0$ after renormalization $p \rightarrow p e^{-\epsilon p}$.

\subsection{Multidimensional interpretation}\label{MI}
\begin{figure}[ht]
 \centering
\includegraphics[width=8.0cm]{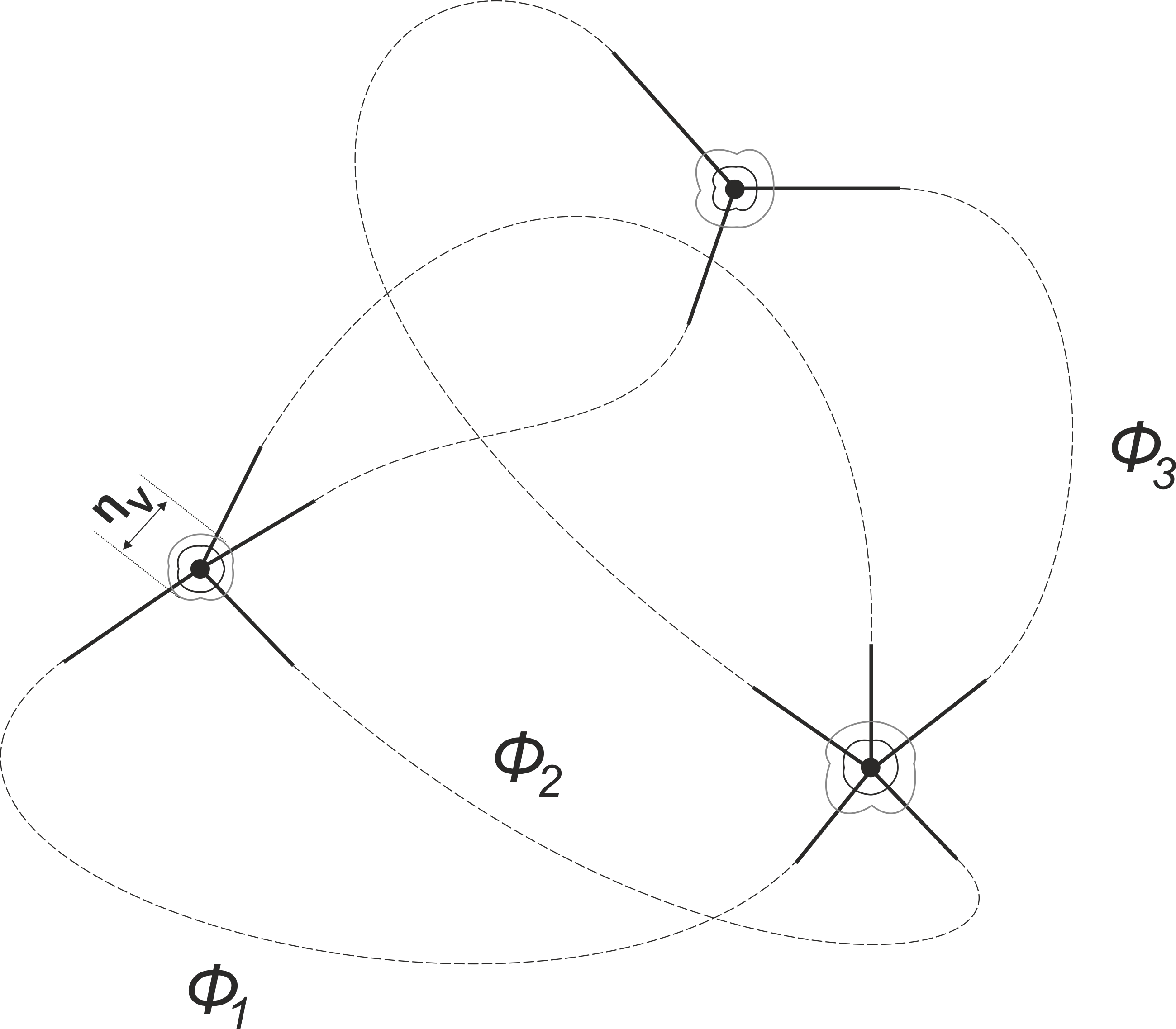}
\caption{Scheme of a number of $6$ fields connected by $3$ particles. The particles have an uncertainty $\eta_{v}$ depending on the velocity $v$ in the orientation of the fields in the space of cognition. The junctions and fields can be pictured as knots and ropes respectively, forming a multidimensional network.}\label{stint}
\end{figure}

As stated at the beginning, the present concept has no fundamental space. The distance $\Omega_I$ is intrinsic to one individual field $I$ and there is a small transition region of order $\eta$ where different fields are connected. These regions are interpreted as elementary particles. The position of one particle related to an individual component of the field is determined by the steep gradient $\frac {\partial \phi_I}{\partial s}$. The parity of the particle is related to the parity of the field components. The individual components of the field, therefore, must be seen as spinors and the particles must be attributed by a 	half-integral spin. From the isomorphism to the 3-dimensional SU(2) symmetry group we may argue that all components can be ordered in a 3-dimensional Euclidean space. This ordering shall only be postulated in a small quasi-local environment around one particle. I will call this mathematical space the `space of cognition', since our cognition orders all physical objects in this space. No assumption about a global space, its topology or dimension has to be made. Figure~\ref{stint} sketches this picture. The quantum-phase-fields form a network where particles are embedded as junctions. Each field is expanded along a 1-dimensional line coordinate and bound by two end points described by gradients of the field. Due to the constraint (\ref{sum}), the coordinates of different field have to be synchronized within the particles of small but finite size $\eta$ along the renormalization condition (\ref{renorm}). The constraint (\ref{sum}) also dictates that there is no `loose end'. The body is closed in itself forming an `universe'.

\section{Generalized Newton laws}\label{GN}
In section \ref{Variational}, the basic relations of the field theoretical framework of the dynamical universe have been set up, based on general consideration and the laws of thermodynamics. They shall now be applied to derive generalized Newton's equation of acceleration and gravitation. Finally a prediction of the structure of the observable universe on ultra long distances will be given.

\subsection{Generalized Newton's equation}\label{GN1}

Let the moving frame of one particle $i$ connect to the field $I$. Then (\ref{acceleration}) yields:
\be\label{Newton2}
U\frac v{c^2}\eta\frac {\partial\phi_I}{\partial s} <w|\frac {\partial}{\partial t}|w>= U\eta \frac {\partial\phi_I}{\partial s} <w|\frac {\partial}{\partial s}|w>.
\ee

Again neglecting quantum effects within the particle, we find by partial integration and using the normalization of the wave function $<w|w>=1$ :
\be\label{Newton3}
\frac {\partial}{\partial t} \frac {U\vec v} {c^2}{\bigm |_{s_i}} =\frac {\partial}{\partial \vec s} U {\bigm |_{s_i}}= \vec f {\bigm |_{s_i}}.
\ee

The position $s_i$ is marked by $\frac{\partial\phi_I}{\partial s} \ne 0$. The direction of the field embedded into the space of cognition defines the directions $\vec s$ and $\vec v$. $\vec f$ is the force acting on the particle by the variation of energy with space. We have Newton's second law. This may be taken as a first prediction of the concept, or simply as a check for consistency.

Inserting the traveling wave solution (\ref{travellingWave}) into the equation of motion (\ref{min}) relates the velocity of the particle to the energy of the field acting on it \footnote{Note that the solution (\ref{travellingWave}) implies $ \\ \eta (1-\frac{{v}^2}{c^2}) \frac {\partial^2}{\partial s^2}\phi  + \frac{\pi^2}\eta (\phi-\frac12)=0$.},
\be\label{velocity}
v = [\frac\partial{\partial s} \phi]^{-1}\frac\partial{\partial t} \phi = \frac {\eta} {\tau} \Delta e \sqrt{1-\frac{v^2}{c^2}}.
\ee

The last equation can be solved for $v$
\be\label{velocity_2}
v = \frac {\eta} {\tau} \frac {\Delta e} {\sqrt{1+ \frac {\eta^2} {{\tau}^2} \frac{{\Delta e}^2}{c^2}}}.
\ee

There is a maximum velocity $v_{max} = c$ which is reached in the limit $\Delta e \rightarrow \infty$.

In expression (\ref{velocity_2}) only the quotient $ \frac {\eta} {\tau}$ appears. We may argue that the time needed to transfer information over the distance $\eta$ is proportional to $\eta$. Then this quotient can be taken as a finite constant in the sharp interface limit $\eta\rightarrow 0$. Since $\tau$ has the dimension of a momentum, it is convenient to use $\tau\propto m_0 c$ with the kinetic constant of dimension of mass $m_0$. Defining a characteristic distance $\bar \Omega$ we set:

\be\label{kinetincConstant}
\frac {\eta} {\tau} = \frac {\bar \Omega} {m_0 c}.
\ee

In this setting and relating the force $\Delta e$ to an absolute energy $\Delta E = \bar \Omega\Delta e$  we see:
\be\label{check}
\lim_{m_0\rightarrow0} |v| = \lim_{m_0\rightarrow0} \frac 1{m_0 c} \frac {|\Delta E|} {\sqrt{1+ \frac{{(\Delta E)}^2}{m_0^2 c^4}}}= c.
\ee

In the limit $m_0\rightarrow 0$ the velocity $|v|$ becomes identical to the constant $c=v_{max}$. Therefore, we can call the maximum velocity the speed of massless particles: speed of light. This result can be seen as the second prediction of the concept, or, again, as a check for consistency with observations and established theories. We end up with the simple relation between the energy of the field and the momentum of its surface states
\be\label{velocity_3}
m_0 v = \frac {\Delta E} {\sqrt{c^2+ \frac{(\Delta E)^2}{m_0^2c^2}}}; \;\; \;\;\frac {m_0 \, c \, v} {\sqrt{1-\frac{v^2}{c^2}}} = \Delta E,
\ee
with the relativistic mass $\frac {m_0} {\sqrt{1-\frac{v^2}{c^2}}}$. No empirical statement about invariance of the speed of light is used. We have, however, the implicit notion of physical space behaving like an `ether', the field, and the upper velocity $c$ corresponds to the well known hyperbolic shock in an elastic medium.

\subsection{Generalized law of gravitation}\label{GG}

A number of $N_i<N$ fields $\phi_I(s,t), I=1...N_i$ connects one single particle $i$ with $N_i$ other particles $j=1...N_i$. For $N_i$ large compared to the dimensionality $D$ of the multi-dimensional space of cognition, it will be impossible that all fields have the same size $\Omega_{ij}\ne \Omega$. We have the generalization of (\ref{EC}) in the reference frame of an individual particle $i$ with $N_i$ attached fields 
\be\label{E_space}
E_i = \sum_{j=1}^{N_i} E_{ij} = -\alpha_i  \sum_{j=1}^{N_i}  \frac {h c}{48 \Omega_{ij}}.
\ee

$E_i$ is the spatial energy of all fields connected to the particle $i$. $E_{ij}=E_J$ is the spatial energy of an individual field $J$ connecting particle $i$ with particle $j$. $\Omega_{ij}$ is the distance between points $i$ and $j$. Balancing the massive energy $U$ with the spatial energy $E_i$ according to the principle of neutrality and defining the characteristic size $\Omega_i~:=~N_i [ \sum_{j=1}^{N_i} \frac1{\Omega_{ij}}]^{-1}$, we evaluate the coupling coefficient for particle $i$ to all other particles $j$:

\be\label{alphaI}
\alpha_i = \frac{96 U \Omega_i}{c N_i h}; \;\;\;\;\;\;\;\; E_i=-\frac{2U{\Omega_i}}{N_i}\sum_{j=1}^{N_i}  \frac 1 {\Omega_{ij}}.
\ee

Up to here, $\Omega_i$ and $\Omega_{ij}$ had been related to the distance between particles. There is, however, no mechanism to evaluate such a distance. At the locus of one particle only the quasi-local spectrum of quantum fluctuations can be used to evaluate the relation between fields and particles. We note that in equation (\ref{alphaI}), only the relative distance $\frac {\Omega_{ij}}{\Omega_i}$ enters. Therefore, we can replace the evaluation of the distances  by the evaluation of the local spectrum of fluctuations acting on the particle $i$. The apparent distance $\tilde\Omega_{ij}$ at position $i$ can be defined from the force $e_{ij}$ by $\tilde\Omega_{ij}=\frac U{e_{ij}}$. Note that in general $\tilde\Omega_{ij} \ne \tilde\Omega_{ji}$, see also discussion in section \ref{speculations}. The characteristic distance $\Omega_i$ will be replaced by the apparent distance $\tilde \Omega_i =  N_i [ \sum_{j=1}^{N_i} \frac{e_{ij}}U]^{-1}$. This apparent distance will be treated in the following as an independent variable acting like a chemical potential equalizing the fluxes of quanta acting on one individual particle from different fields.

We find by partial integration, in analogy to (\ref{Newton2}) and (\ref{Newton3}), the force $\vec f_{ij}$ acting on particle $i$ from the fields connecting it to particle $j$ in accordance to Newton's first law:
\be\nonumber
 U \frac {\vec v_i}{c^2}\sum_{j=1}^{N_i} \left[\frac {\partial\phi_{ij}}{\partial s} <w|\frac {\partial}{\partial t}|w>\right]=\sum_{j=1}^{N_i}\left[E_{ij} \frac {\partial\phi_{ij}}{\partial s} <w|\frac {\partial}{\partial s}|w>\right],
\ee
\be\label{Newton4}\frac\partial{\partial t}\frac {U \vec v_i}{c^2}|_{s_i}= - \sum_{j=1}^{N_i} \vec n_{ij}\frac\partial{\partial s} E_{ij}|_{s_i} =\sum_{j=1}^{N_i} \vec f_{ij}|_{s_i} ,
\ee

where $\vec n_{ij}$ is the normal vector of the field in the space of cognition, evaluated at the position of particle $i$.

The energy has two distance dependencies: the dependence of $\alpha_i$ on $\tilde \Omega_i$ and the dependence of the energy of the individual field $E_{ij}$ on the apparent distance $\tilde \Omega_{ij}$. Both must be varied independently. One finds the force $\vec f_{ij}$ of particle $j$ acting on particle $i$, using (\ref{alphaI})
\bea\label{F_semifinal}
\nonumber \vec f_{ij} &=& - \vec n_{ij} [\frac d{d \tilde \Omega_{ij}} E_i|_{\tilde \Omega_i=const}+  \frac d{d \tilde \Omega_i} E_i|_{\tilde\Omega_{ij}=const}]\\
&=& \vec n_{ij} \frac {2 U \tilde\Omega_i}{N_i \tilde\Omega_{ij}^2}(1 - \frac{\tilde\Omega_{ij}} {\tilde\Omega_i})
\eea
and the total force $\vec f_{i}$  from all masses $j$
\be\label{F_total}
 \vec f_i = \frac {2 U \tilde\Omega_i}{N_i}\sum_{j=1}^{N}  \frac {\vec n_{ij}}{\tilde\Omega_{ij}^2}(1 - \frac{\tilde\Omega_{ij}} {\tilde\Omega_i}).
\ee

$\tilde\Omega_i$ defines a marginal distance where the force vanishes. Massive particles $i$ and $j$, which are distant by $\tilde\Omega_{ij} < \tilde\Omega_i$, are accelerated towards each other by the quantum fluctuations they receive from all other particles, behaving like being attracted. The masses $i$ and $j$, which are distant by $\tilde\Omega_{ij} > \tilde\Omega_i$, repel each other.  $\tilde\Omega_i$ separates interactions from attractive to repulsive. Thereby, a cloud of $N_i$ fields connecting $N_i+1$ massive particles separates spontaneously into dense and dilute regions. In the limit $\tilde\Omega_{ij} \ll \tilde\Omega_i$, (\ref{F_total}) reduces to the classical Newton law of gravitation if we identify the pre-factor $\frac {2 U \tilde\Omega_i}{N_i}$ divided by the product of the masses $m_i$ and $m_j$ with the coefficient of gravitation in the local environment of elementary mass $i$, $G_i$:

\be\label{F_final}
 \vec f_i = G_i\sum_{j=1}^{N} \vec n_{ij} \frac {m_i m_j }{\tilde\Omega_{ij}^2}(1 - \frac{\tilde\Omega_{ij}} {\tilde\Omega_i}).
\ee

This is the final prediction of the concept. It is a mere result of quantum fluctuations in finite space and the postulate of energy conservation in the strong (quasi-local) form. The generalized law of gravitation (\ref{F_final}) predicts repulsion of distant masses. This repulsion will increase further unbound in distance. This statement offers an explanation of the observed acceleration of expansion of the universe \cite{Riess1998}.

\subsection{Size of the voids in the universe}\label{voids}
In order to derive an estimate of the marginal, or characteristic distance $\tilde\Omega_i$, I assume that hydrogen and neutrons are the dominant elements in the observable universe. Taking the mass of the universe $M\approx~10^{52}~kg$  \cite{Persinger2009} with the mass of the hydrogen atom $m_h \approx u\approx 1.66~10^{-27} kg$ and the mass of the neutron $m_n\approx u$, we find the number of masses visible from the earth $N_E$ and the characteristic distance $\tilde \Omega_E$ based on the measured gravitational coefficient on earth $G_E\approx 6.67 \, 10^{-11} \frac {m^3}{kg s^2}$
\be\label{N}
N_E = \frac M {u}; \;\;\;\;\;\;\;\;  \tilde \Omega_E = \frac{G_E M}{2 c^2} \approx  10^{24} m.
\ee

This numerical value of $\tilde\Omega_E$ corresponds well to the size of the so-called `voids' \cite{Mueller2000}. The voids are regions in the universe which are nearly empty of masses: masses at the rim of one void repel each other so that no mass enters one void by `gravitational' forces.

\section{Discussion and interpretation}\label{speculations}

In the previous section a rigorous derivation has been presented from which generalized Newton's equations, invariance of speed of light and repulsive gravitational action on ultra-long distances are derived. The latter is, of course, consistent with Einstein's equation with a finite cosmological constant, though the approach is fundamentally different. The question is how to `adjust' such a cosmological constant, see \cite{Hebecker2000}. In the present concept there is no `global' constant. The marginal length is formulated from a quasi-local energy balance. Let me explain this in more detail: As stated in the beginning, there is no fundamental, absolute space, neither 1-dimensional nor multi-dimensional. Space is defined by the (negative) energy content of the volume of the quantum-phase-fields $\phi_I \equiv 1$ on the one hand. Within one particle $\phi_I < 1$, $\frac{\partial \phi_I}{\partial s} \ne 0$, on the other hand, it is related to a 1-dimensional metric which distinguishes different values of the field. The particles have a small but finite size $\eta$ where several fields coincide. Here, the wave functions of different fields have to be superposed non-locally. Outside the particle, the wave function collapses into a single field wave function, which carries, however, the probabilistic quantum information of the particle to the particle at the opposite end of the of the field. The nonlocal region of the particle hereby may be extremely small, as discussed by Zurek \cite{Zurek2001}. The expression `quasi-local' shall emphasize that we have a non-local theory with highly localized quantum states. A detailed quantum mechanical description is far beyond the scope of this work. We might, however, relate the existence of separated volume regions of the fields to `hidden variables' in Bohm's interpretation of quantum mechanics \cite{Bohm1952a, Bohm1952b}: individual energy quanta, emitted from one particle into the volume of one field, already `know' the particle where they will be received, since one field component connects two distinct particles only. The quantum-statistical process of where to emit to, is attributed to the particles only. This interpretation of the exchange mechanism may also be related to Wheeler-Feynman's absorber theory of light \cite{Wheeler1945}, or Cramer's transactional interpretation of quantum mechanics \cite{Cramer1986}, which connects the emission of a light quantum to an unique future event of absorption. I leave closer interpretation to future work. Within the quasi-local region of one particle, an 'action at a distance' in the sense of the EPR paradox exists: entangled quantum states (for a recent discussion of the EPR paradox see \cite{Kiefer2015}, in German). The particles exchange energy with the field by an exchange flux for which a continuity equation in the classical sense must hold: generalized Newton's equation (\ref{Newton4}). It is hereby unnecessary to `know' the actual energy content of any state of the body of the universe, except the homogeneous initial state (without space, time and energy). Any future state must have the same energy if no energy is created or destroyed. The mechanism of transferring action between massive bodies in the present concept is: emitting quanta into the field, or receiving quanta from the field. This happens in the quasi-local environment of one individual particle. According to the definition of space by the spectrum of quantum fluctuations, de- and increasing the spectrum of fluctuation means contraction and elongation of space respectively. It is evident, however, that this change of length will not happen instantaneously. There will be fluctuations of quanta within the field which I assume to dissipate with the speed of light. In other words, action between bodies is transferred with the speed of light. There will be a `delay' of action. We might argue that the dependence of the apparent size of a field in the case of accelerated particles $\Omega_{ij} \ne \Omega_{ji}$ on the position of the observer and the direction of acceleration is complementary to the gravitational time dilatation in general relativity \cite{Einstein1916}. Here more detailed investigations are necessary in future work, too.

In light of the present concept, `dark matter' looses its mystery. We  simply relax the idea that all particles and fields have to be connected to all others. Particles which do not have a connecting field to the observer are `invisible' since there is no space through which the light could travel. But they will be detectable by their influence on particles which they are connected to and which have a direct connection to the local observer.

Finally, let me try an estimate of the size $\eta$ of one particle. Comparing the energetic and spatial constants of the expressions for the volume of a field (\ref{energydensity}) and (\ref{EC}) we read, using the numerical value of the number of particles in the observable universe $N_E$ from (\ref{N}),

\be\label{eta_single}
\eta_\text{single} \propto \frac {\alpha h c}U \approx \frac {\tilde \Omega_E}{N_E} \approx 10^{-55}m.
\ee

The proportionality constant is of order $1$ depending on the volume integration over the particle which is done here only for the special case of two connecting fields. Despite the large uncertainties in several ingredients to determine the actual value of $\eta$, it can be concluded that the size of a junction which relates to one single elementary particle $\eta_{single}$, like a neutrino, must be considered as `point-like', far below Planck's length. A triplet of three quarks in a confined state, however, defines a 2-dimensional object. This means that the number of connected fields $N_E$ in (\ref{eta_single}) must be related to an area proportional to $\eta^2$. The radius of this area $\eta_{triple}$ is estimated to be

\be\label{eta_triple}
\eta_\text{triple} \approx \frac {\tilde \Omega_E} {\sqrt{N_E}} \approx 10^{-16}m,
\ee
comparable to the size of a neutron.

\section{Conclusion}\label{conclusion}

The new monistic concept of matter treats the energy, ordered by a set of quantum-phase-fields, as the only existing substance. The dual elements of matter, mass and space, are described by volume- and gradient-energy contributions of the fields respectively. The concept is based on the statement that energy can neither be created nor destroyed, the first law of thermodynamics. The origin of the universe is treated as a spontaneous decomposition of the symmetric state of $0$ energy (`nothing') into `matter', mass and space, by the demand of entropy production, the second law of thermodynamics. The time evolution of the fields dictates the time dependence of the Hamiltonian and the wave function. The wave function $|w>$ is decomposed into single component wave functions $|w_I>$ in the limiting case of quasi-stationary fields and constructed explicitly. Space is attributed with negative energy, mass is attributed with positive energy. The physical space is a one-dimensional box between two elementary particles forming the endpoints of space. Quantum fluctuations in finite space with discrete spectrum, compared to a continuous spectrum, define the negative energy of space. The junctions between individual components of the field define elementary particles with positive energy.  The energy of mass is the condensation of those fluctuations which do not fit into finite space. Comparison of the energy of mass to the energy of space defines the coupling coefficient $G_i$ between an individual elementary particle $i$ and the spaces it is embedded in. It depends on the position of one elementary mass $i$ in space and time relative to all other masses. By varying the energy of space with respect to distance, the action on the state of masses is derived. This leads to a generalized law of gravitation which shows attractive action for close masses and repulsive action for masses more distant than a marginal distance $\tilde \Omega_E$. This distance is correlated to the size of the largest structures in the universe observed in the reference frame of our solar system. The predicted marginal length $\tilde \Omega_E$ correlates well with the observed size of the voids in the universe.

It must be stated clearly that the new quantum-phase-field concept is not a priori in conflict with general relativity since it has no restriction concerning the topology of a global multi-dimensional space of cognition. The new contribution of the concept is the quasi-local mechanism of balancing in- and out-going quantum fluctuations on the field at the position of the observer. The concept sticks strictly to the demand of energy conservation. It makes a prediction for gravitational action on ultra-long distances. This prediction can be verified experimentally by investigating trajectories of large structures in the universe. The presented concept might open a door towards a new perception of physics where thermodynamics, quantum mechanics and cosmology combine naturally.


\section*{Acknowledgement} The author would like to thank Claus Kiefer, Cologne, for helpful suggestions and discussions;  Dmitri Medvedev, Bochum/Novosibirsk, for providing the velocity dependent traveling wave solution; Friedrich Hehl, Cologne, for revealing some inconsistencies in the original manuscript and grounding him to reality; Fathollah Varnik, Bochum, for critical reading of the manuscript. 

\newpage
\section{References}

\bibliographystyle{unsrt}

\end{document}